\documentclass[12pt]{article}
\DeclareUnicodeCharacter{0301}{\'{e}}
\usepackage{amssymb}
\usepackage{amsmath}
\usepackage{amsthm}
\usepackage{color}
\usepackage{comment}
\usepackage{enumitem}		
\usepackage{geometry}		
\usepackage{graphicx}
\usepackage{authblk}
\usepackage{amssymb}
\usepackage{listings}
\usepackage{xcolor}
\usepackage{caption}
\usepackage{color}
\usepackage{algorithm}
\usepackage{algpseudocode}

\usepackage{subcaption}

\makeatletter
	\@addtoreset{equation}{section}
\makeatother

\let\originalleft\left
\let\originalright\right
\renewcommand{\left}{\mathopen{}\mathclose\bgroup\originalleft}
\renewcommand{\right}{\aftergroup\egroup\originalright}

\newlist{romanlist}{enumerate}{3}
\setlist[romanlist]{label=\roman*),ref=(\roman*)}

\begin{document}

\newcommand{\cF}{\mathcal{F}}
\newcommand{\cP}{\mathcal{P}}
\newcommand{\cR}{\mathcal{R}}
\newcommand{\cS}{\mathcal{S}}
\newcommand{\cT}{\mathcal{T}}
\newcommand{\ee}{\varepsilon}
\newcommand{\rD}{{\rm D}}
\newcommand{\re}{{\rm e}}

\newtheorem{theorem}{Theorem}[section]
\newtheorem{corollary}[theorem]{Corollary}
\newtheorem{lemma}[theorem]{Lemma}
\newtheorem{proposition}[theorem]{Proposition}

\theoremstyle{definition}
\newtheorem{definition}{Definition}[section]


\title{Torus and hyperchaos in 3D Lotka-Volterra map}

\author[1]{Sishu Shankar Muni}

\affil[1]{School of Digital Sciences,\\ Digital University Kerala\\
Thiruvananthapuram, PIN 695317, Kerala, India \newline
E-mail: sishushankarmuni@gmail.com}

\maketitle


\begin{abstract}
In this study, we investigate the occurrence of a three-frequency quasiperiodic torus in a three-dimensional Lotka-Volterra map. Our analysis extends to the observation of a doubling bifurcation of a closed invariant curve, leading to a subsequent transition into a state of hyperchaos. The absorption of various saddle periodic orbits into the hyperchaotic attractor is demonstrated through distance computation, and we explore the dimensionality of both stable and unstable manifolds. Various routes to  cyclic and disjoint quasiperiodic structures are presented. Specifically we showcase the transition from a saddle-node connection to a saddle-focus connection, leading to the formation of quasiperiodic closed cyclic disjoint curves, as revealed \textcolor{black}{by} the computation of one-dimensional unstable manifold. Additionally, we show an unusual transition from a period-two orbit to a period-six orbit and uncover the mechanism related to two subsequent bifurcations: a) subcritical Neimark-Sacker bifurcation, and (b) saddle-node bifurcation. Our approach involves the use of computational methods for constructing one-dimensional manifolds, extending saddle periodic orbits through a one-parameter continuation, and employing a multi-dimensional Newton-Raphson approach for pinpointing the saddle periodic orbits in the three-dimensional map. 
\end{abstract}
\section{Introduction}
\label{sec:intro}

A logistic model is a mathematical model which describes \textcolor{black}{the} dynamics of a population over time. It is \textcolor{black}{modeled} by a system of differential equations. An Euler discretization of the differential equation based logistic model  gives a logistic map \cite{May1976}. The logistic map is considered as a simplest one-dimensional map which is considered to understand many of the bifurcation phenomena to date.  It has found its applications in several areas of science, engineering, ecological applications \cite{Kawano2020}.  There has been maps developed to understand population dynamics like the Ricker map \cite{BravodelaParra2013}, Lotka-Volterra map \cite{Bischi2010,Blackmore2001}, Beverton-Holt model \cite{ChChaoui2022}. Moreover nonlinear dynamic techniques have been very much essential to uncover many new \textcolor{black}{phenomena} in biological models of neurons \cite{Muni23a,ZS22AEU,ZSChaos22,Chialvo22,IZH22}, atherosclerosis \cite{Chalmers2015}, mild atherosclerosis \cite{Mukherjee2024}, population dynamics \cite{Adachi2022}, and many more. The latter maps are well studied in terms of routes to chaos, dynamics, and bifurcations. The Lotka-Volterra equations, in their continuous form, have been foundational in ecological modeling by describing the dynamics between predator and prey populations in continuous time. The extension to a discrete framework introduces a time-stepping mechanism, allowing for a more realistic portrayal of ecological phenomena influenced by discrete events, such as reproduction and predation events occurring at discrete intervals. This transition from continuous to discrete dynamics is especially pertinent in ecological contexts, where events unfold in distinct, quantifiable steps.

The three-dimensional nature of the model accommodates the consideration of additional factors, such as competing species or mutualistic relationships, expanding its applicability beyond traditional predator-prey scenarios. By incorporating additional dimensions, the model can elucidate complex ecological dynamics, including coexistence, oscillations, and stability, in more realistic and nuanced ecological scenarios.

The discrete Lotka-Volterra map offers a unique perspective on the role of discrete dynamics in shaping ecological communities. Doubling bifurcation is prevalent in many biological, \textcolor{black}{and} mechanical systems like in the Hindmarsh-Rose neuron \cite{Muni23a}, \cite{Muni2024}, \textcolor{black}{and} mechanical systems \cite{Vibr15}. We show that Lotka-Volterra map, a model for population dynamics also showcases the doubling bifurcation of closed invariant curves. Additionally, we show that this biological model also showcases a three-frequency quasiperiodic torus. This makes the Lotka-Volterra map a candidate for further exploration of the bifurcation for the three-frequency resonant torus.

Typically, ecological populations exhibit non-overlapping generations, as observed in insects such as gypsy moths. In instances where discrete time is more suitable for modeling, methods such as the Euler forward scheme or semi-discretization scheme are commonly employed \cite{Ma2023}. In many instances involving discrete maps, one can identify a confined and limited subset that remains unchanged under the influence of the map. These subsets are of particular interest within the context of dynamical systems, as they provide insights into the behavior of trajectories for all points within the set in the phase space. In a study by Balibrea et al. \cite{Balibrea2006,LuisGarcaGuirao2008}, the authors demonstrated the transitive nature of the Lotka-Volterra map. Moreover, they inferred that the Lotka-Volterra map exhibits the property of mixing.

Moreover, we show a peculiar transition from a period-two orbit to a period-six orbit and the mechanism behind \textcolor{black}{such a transition}. After computation and continuation of saddle periodic orbits, we show that such a transition follows two steps comprising of 1) a subcritical Neimark-Sacker bifurcation of a stable period-two orbit leading to the formation of a repelling quasiperiodic closed invariant curve, and 2) a saddle-node bifurcation of the repelling closed invariant curve leading to the formation of a stable period-six orbit. We further showcase that the 3D map under consideration develops hyperchaos with two positive Lyapunov exponents in a wide range of parameter space. Nonlinear systems exhibiting hyperchaos are important from \textcolor{black}{the} viewpoint of real-world applications. Such systems find applications in cryptography, encryption schemes \cite{Yang2020}, and secure communication schemes \cite{KOCAREV1992}.

Through rigorous mathematical analysis and numerical simulations, we aim to unravel the behavior of multi-species interactions, providing insights into the stability and resilience of ecological systems under discrete influences. Furthermore, our exploration extends beyond the basic understanding of ecological dynamics to investigate the impact of key parameters, initial conditions, and bifurcations on the system's behavior.

The main contributions of the paper are listed as follows:
\begin{itemize}

    \item Prevalence of three-frequency resonant torus via a quasiperiodic Hopf \textcolor{black}{bifurcation}.
    \item Doubling bifurcation of quasiperiodic closed invariant curve.
    \item Unfolding a peculiar transition of a stable period-two orbit to a stable period-six orbit.
    \item Various routes to hyperchaos and absorption of saddle periodic orbits.
    \item Route to the formation of disjoint cyclic closed invariant curves and the interplay of saddle-node and saddle-focus connections.

\end{itemize}
The paper is organized as follows: In \S \ref{sec: Model}, the three-dimensional discrete Lotka-Volterra map is introduced. In \S \ref{sec:twoparam}, a two-parameter Lyapunov chart is \textcolor{black}{analyzed}. In \S \ref{sec:quasibif}, a route to the formation of \textcolor{black}{the} three-frequency torus is shown via a quasiperiodic Hopf bifurcation. In \S \ref{sec:doubling}, quasiperiodic torus doubling bifurcation is discussed. In \S \ref{sec:doubling}, torus doubling bifurcation of closed invariant curve and route to hyperchaos is discussed. In \S \ref{sec:transition}, an unusual transition from period-two orbit to period-six orbit is shown and a mechanism is deduced. The paper concludes by providing a summary, future directions, and conclusions. 

\section{Three-dimensional Lotka-Volterra map}
\label{sec: Model}
Let us consider the three-dimensional discrete Lotka-Volterra map \textcolor{black}{modeled} for studying the population dynamics in case of three different species \cite{Bischi2010,Blackmore2001}.  The three-dimensional mapping is given as follows  
\begin{equation}
    \begin{aligned}
    x_{n+1} &= x_{n} + Rx_{n}(1-x_{n} - \alpha y_{n} - \beta z_{n}),\\
    y_{n+1} &= y_{n} + Ry_{n}(1- \beta x_{n} - y_{n} - \alpha z_{n}),\\
    z_{n+1} &= z_{n} + Rz_{n}(1-\alpha x_{n} - \beta y_{n} - z_{n}),
    \end{aligned}
    \label{eq:STMmap}
\end{equation}
where $x,y,z$ \textcolor{black}{represent} the \textcolor{black}{three species of population},  $\alpha, \beta$ \textcolor{black}{represent the parameters that describe different kinds of interactions between the three species of populations $x,y,z$, and $R$ represent the birth-rate parameter}. A detailed fixed point analysis is carried out in \cite{Balibrea2006} but to our knowledge, no remarks \textcolor{black}{are} made on the invertibility of the map \eqref{eq:STMmap}. It is important to understand if the map under consideration is invertible as \textcolor{black}{the} non-invertibility of the map explains the stretching and folding action. Let us consider $R=1,\alpha=1,\beta=1$, then note that both $(0,0,0)$ and $(1,1,1)$ map to $(0,0,0)$  implying that the three-dimensional discrete Lotka-Volterra map is non-invertible. 

We start our analysis with a two-parameter Lyapunov chart in the following section \textcolor{black}{which is helpful to identify different types of bifurcations.}
\section{Two parameter Lyapunov chart}
\label{sec:twoparam}
\begin{figure*}[!htbp]
\centering
\includegraphics[width=0.7\textwidth]{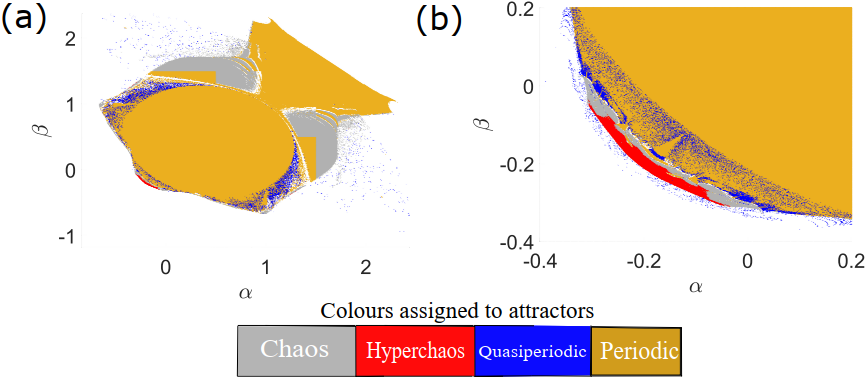}
\caption{In (a), a two-parameter Lyapunov chart of \eqref{eq:STMmap} in the $\alpha-\beta$ plane with $R=1.4$. Orange color denotes the periodic or fixed point regime. Grey denotes the chaotic regime. Blue denotes the quasiperiodic regime, and red denotes the hyperchaotic regime with two positive Lyapunov exponents. A zoomed version of (a) is shown in (b).}
\label{fig:LyapunovChartThreeLV}
\end{figure*}

 In Fig. \ref{fig:LyapunovChartThreeLV}, we show a two-parameter $\alpha-\beta$ Lyapunov chart. Other combinations of \textcolor{black}{parameters} reveal similar charts and hence are not shown.  Different colors are allotted for the prevalence of attractors at those parameter regimes, see the legend. \textcolor{black}{The orange} color denotes the periodic regime or fixed point regime. Blue \textcolor{black}{color} denotes \textcolor{black}{the} quasiperiodic regime, grey \textcolor{black}{color} denotes the chaotic regime, and red \textcolor{black}{color} denotes the hyperchaotic regime with at least two positive Lyapunov exponents. \textcolor{black}{To} understand the route to hyperchaos, a zoomed version of \textcolor{black}{the} Lyapunov chart in (a) is shown in Fig. \ref{fig:LyapunovChartThreeLV} (b).   It is interesting to note that the shape of the Lyapunov chart appears to be symmetric and is an important aspect to be explored further. \textcolor{black}{The symmetry of Lyapunov exponents in one-dimensional bifurcation structures in one-dimensional maps has been discussed in \cite{Guchi16}. It has been shown that such a symmetry is due to the prevalence of an invariant transformation from paired parameter values to another set of parameter values and as a consequence the type of attractors remain the same. We believe such techniques can also be used in the three-dimensional Lotka-Volterra map under consideration. To understand the overall symmetry in the two-dimensional Lyapunov chart, various one-parameter bifurcation diagrams can be constructed to verify the symmetry of Lyapunov exponents. Finally, a transformation can be sought to understand the mechanism behind the symmetrical structure. }

\section{Quasiperiodic Hopf bifurcation}
\label{sec:quasibif}
\begin{figure*}[!htbp]
\centering
\includegraphics[width=0.9\textwidth]{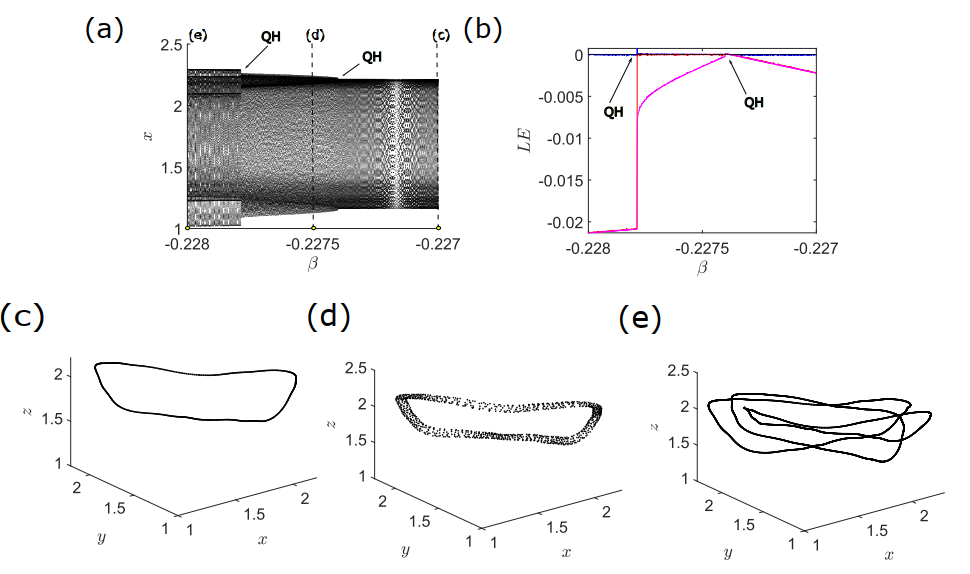}
\caption{Occurrence of quasiperiodic Hopf bifurcation. In (a), a one-parameter bifurcation diagram is shown. In (b), a one-parameter Lyapunov exponent spectrum is shown. In (c), (d), (e), phase portraits for a quasiperiodic closed invariant curve, a three-frequency torus, and a three disjoint cyclic closed invariant curve is shown.}
\label{fig:CollageThreeFreqTorus}
\end{figure*}

In Fig.  \ref{fig:CollageThreeFreqTorus} (a), a one-parameter bifurcation diagram of $x$ vs $\beta$ is considered. A corresponding Lyapunov exponent spectrum diagram is considered in Fig. \ref{fig:CollageThreeFreqTorus} (b). Formation of a higher frequency quasiperiodic torus from a lower frequency torus occurs through a quasiperiodic Hopf bifurcation. We illustrate the occurence of quasiperiodic Hopf bifurcation in this case. Notice the change in the structure of \textcolor{black}{the} quasiperiodic regime as parameter $\beta$ is decreased from $-0.227$, see Fig. \ref{fig:CollageThreeFreqTorus}. A similar behavior can also be observed in the spectrum of Lyapunov exponents, see Fig. \ref{fig:CollageThreeFreqTorus} (b). A quasiperiodic Hopf bifurcation leads to the formation of a higher frequency quasiperiodic torus. Three different cross-sections are chosen to illustrate the transformation of a closed quasiperiodic invariant curve to a three-frequency torus, see Fig. \ref{fig:CollageThreeFreqTorus} (c) - (e).    

For $-0.2273 < \beta < -0.227$, we observe a three cyclic quasiperiodic closed invariant curve, see Fig. \ref{fig:CollageThreeFreqTorus} (c). When $\beta < -0.2273$, a quasiperiodic-Hopf bifurcation takes place leading to the formation of a three-frequency quasiperiodic torus, see Fig. \ref{fig:CollageThreeFreqTorus} (d).

 A three cyclic disjoint closed invariant curve (see Fig. \ref{fig:CollageThreeFreqTorus} (e)) bifurcates to a three-frequency torus in Fig. \ref{fig:CollageThreeFreqTorus}(d). 
 A similar quasiperiodic Hopf bifurcation takes place when $\beta$ is increased from $-0.228$. A single quasiperiodic closed invariant curve bifurcates to a three frequency torus. In each of the quasiperiodic Hopf bifurcation, two Lyapunov exponents are zero which confirms the formation of a three frequency quasiperiodic torus. 

\section{Quasiperiodic doubling bifurcation}
\label{sec:doubling}
\begin{figure*}[!htbp]
\centering
\includegraphics[width=0.9\textwidth]{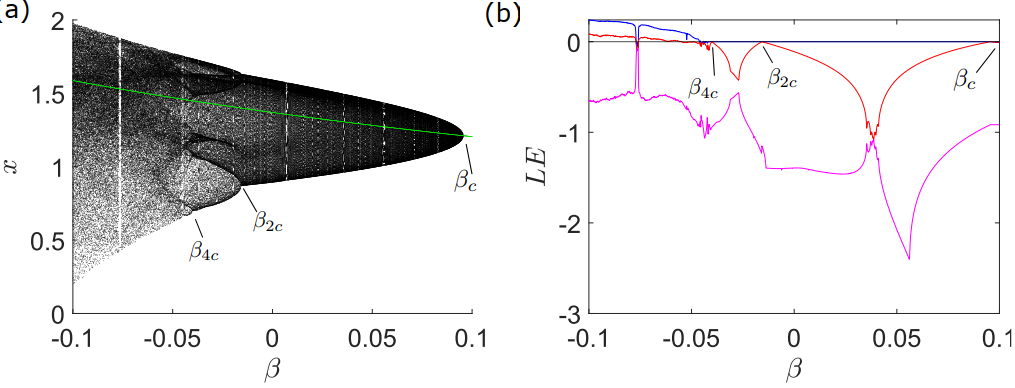}
\caption{A one-parameter bifurcation diagram of $x$ vs $\beta$ along with the continuation of saddle fixed point (in green). In (b), its corresponding one-parameter Lyapunov exponent spectrum with respect to parameter $\beta$. }
\label{fig:CollageOneParamDoubling}
\end{figure*}

\begin{figure*}[!htbp]
\centering
\includegraphics[width=0.9\textwidth]{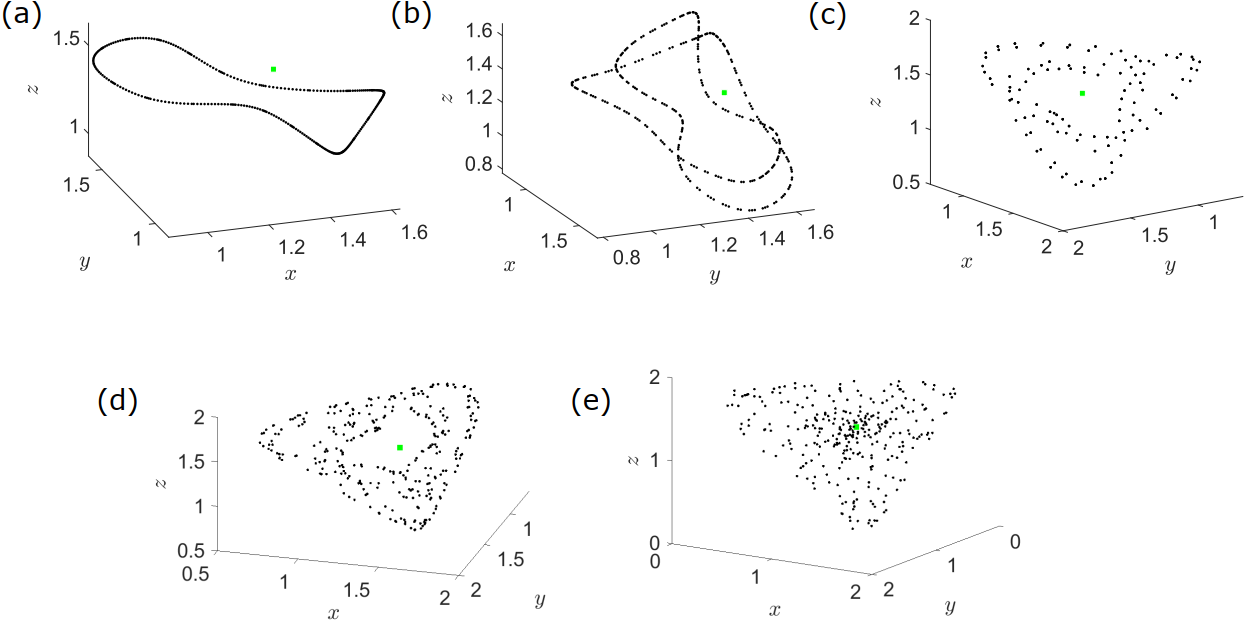}
\caption{Coexistence of the attractors (in black) and saddle fixed point (in green squares). In (a), coexistence of closed invariant curve and saddle fixed point in green for $\beta = 0.08729$. In (b), coexistence of length-doubled closed invariant curve and saddle fixed point for $\beta = 0.0742971$. In (c), further length-doubled closed invariant curve and saddle fixed point for $\beta = 0.06592$. In (d), formation of chaotic attractor for $\beta = 0.0609093$. In (e), formation of hyperchaotic attractor for $\beta = 0.0630183$. Other parameters $R = 1.4, \alpha = -0.227$ are fixed. }
\label{fig:CollagePhaseSpace_Doubling}
\end{figure*}

A one-parameter bifurcation diagram of $x$ vs $\beta$ is shown in Fig. \ref{fig:CollageOneParamDoubling} (a). A stable fixed point undergoes a super critical Neimark-Sacker bifurcation at $\beta = \beta_{c}$ leading to the formation of a closed invariant curve and a pair of eigenvalues of the stable fixed point crosses the unit circle rendering it a saddle fixed point which continues to exist.
The saddle fixed point is continuated further as shown in green dots. As $\beta$ is further decreased, the quasiperiodic closed invariant curve undergoes a doubling bifurcation at $\beta = \beta_{2c}$ leading to the formation of a length doubled quasiperiodic closed invariant curve, see Fig. \ref{fig:CollagePhaseSpace_Doubling} (b). This is also detected in the Lyapunov spectrum in which a second Lyapunov exponent touches zero. A subsequent doubling bifurcation occurs at $\beta = \beta_{4c}$, see Fig. \ref{fig:CollagePhaseSpace_Doubling} (c). As $\beta$ decreases further, formation of chaotic attractor takes place, see Fig. \ref{fig:CollagePhaseSpace_Doubling} (d). Notice that the saddle fixed point is not yet abosrbed in the chaotic attractor. As $\beta$ further decreases, a Lyapunov exponent smoothly crosses zero and becomes positive, leading to the formation of a hyperchaotic attractor. Notice also that the saddle orbit has been absorbed in the hyperchaotic attractor, see Fig. \ref{fig:CollagePhaseSpace_Doubling} (e). The saddle fixed point, denoted in green dots is completely lying inside the hyperchaotic attractor.  

\begin{figure}[!htbp]
\centering
\includegraphics[width=0.7\textwidth]{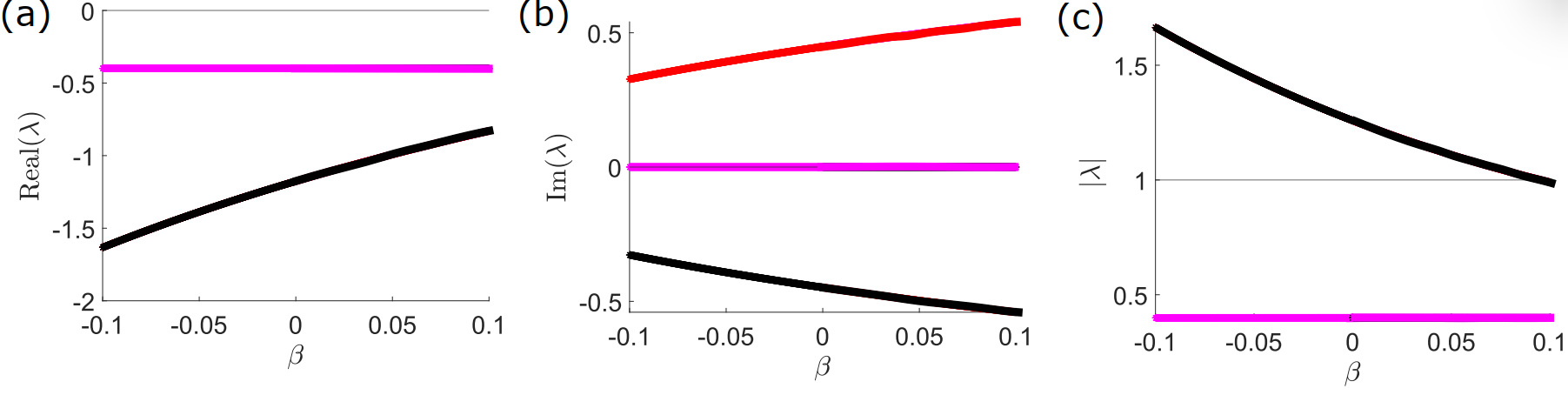}
\caption{Variation of the eigenvalues of the saddle fixed point with respect to parameter $\beta$. In (a), variation of real part of the eigenvalue with respect to parameter $\beta$ is considered. In (b), variation of imaginary part of the eigenvalue with respect to parameter $\beta$. In (c), variation of magnitude of the eigenvalue with respect to parameter $\beta$.}
\label{fig:CollageEigenDoubling}
\end{figure}  

It is also important to understand the dimensionality of the saddle periodic orbits absorbed inside the hyperchaotic attractor. From the continuation of eigenvalues with respect to parameter $\beta$ in Fig. \ref{fig:CollageEigenDoubling} (a) - (c), we can observe that after the first Neimark-Sacker bifurcation, two eigenvalues attain values greater than one. With further decrease in $\beta$, we see that the number of eigenvalues with absolute value greater than one is equal to two and thus the number of unstable directions of the saddle fixed point is two implying an expansion in two directions. Observe that the real part of the eigenvalues are negative, see Fig. \ref{fig:CollageEigenDoubling} (a). The only condition that needs to be satisfied is the homoclinic condition that is a transversal intersection of the stable and unstable manifold of the saddle fixed point in order for the hyperchaotic attractor to be classified as Shilnikov hyperchaotic attractor. This would require sophisticated computation of one-dimensional stable manifold of the Lotka-Volterra map, a non-invertible map.  It would be an interesting problem for future work. 

\section{Transition from period-2 to period-6 orbit}
\label{sec:transition}  
\textcolor{black}{In this section we discuss a direct transition from a period-two orbit to a period-six orbit which is interesting. As it is well known a period-doubling route is more prevalent.} To understand the mechanism behind this tripling, we dive deeper into the role played by the saddle periodic orbits. 

\begin{figure*}[!htbp]
\centering
\includegraphics[width=0.9\textwidth]{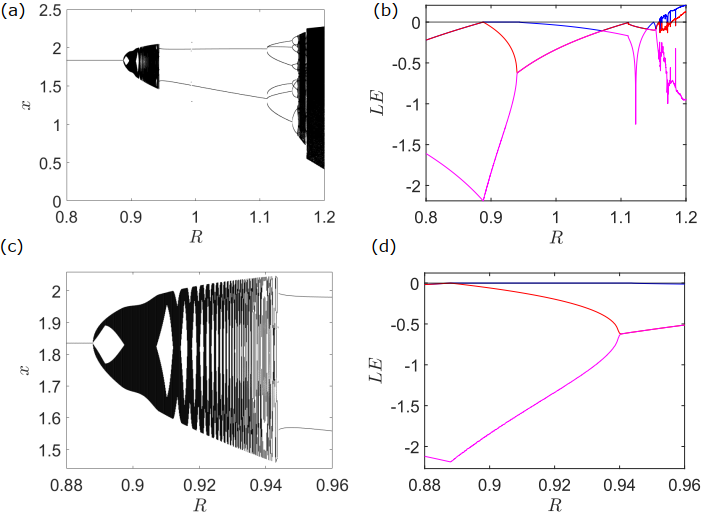}
\caption{In (a), a one-parameter bifurcation diagram of $x$ vs $R$. In (b), corresponding Lyapunov spectrum is shown with respect to $R$.  Parameters are fixed as $\alpha = -0.227, \beta = -0.228$.}
\label{fig:CollageOneParam_Hyperchaos}
\end{figure*}

With the variation of parameter $R$, see Fig. \ref{fig:CollageOneParam_Hyperchaos} (a)-(b), we observe that a stable fixed point undergoes a supercritical Neimark-Sacker bifurcation leading to the formation of quasiperiodic closed invariant curve. \textcolor{black}{In Fig. \ref{fig:CollageOneParam_Hyperchaos} (c), a zoomed version of one-parameter bifurcation diagram is presented and the corresponding Lyapunov spectrum is shown in Fig. \ref{fig:CollageOneParam_Hyperchaos} (d). We note that during the occurrence of Neimark-Sacker bifurcation at $R = 0.887$, one of the Lyapunov exponent reaches zero implying the formation of the closed invariant curve. One of the Lyapunov exponent remains at zero till the occurrence of the saddle-node bifurcation of the closed invariant curve at $R = 0.943$, leading to the formation of mode-locked period-two orbit. Further as $R$ increases, the Lyapunov exponent becomes negative implying the existence of regular periodic behavior.}

\begin{figure*}[!htbp]
\centering
\includegraphics[width=0.9\textwidth]{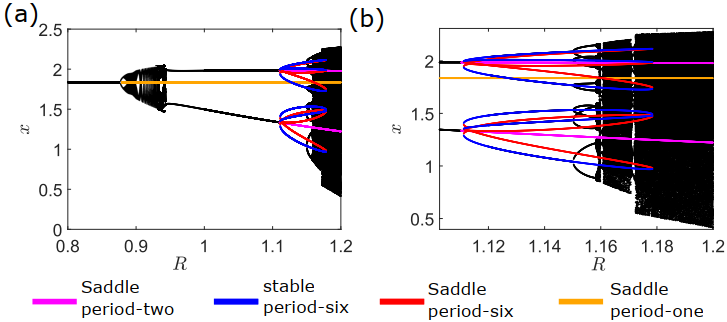}
\caption{A detailed one-parameter bifurcation diagram of $x$ vs $R$. In (a), saddle period-one orbit, period-two orbit, period-six orbit are continuated. In (b), a zoomed part of (a) is presented. }
\label{fig:HyperchaosVaryR}
\end{figure*}

With further increase in parameter $R$, we observe occurrence of a saddle-node bifurcation leading to the formation of a period-two orbit. With further increase in parameter $R$, the period-two orbit undergoes a transition to a period-six orbit. Such type of transition are common in maps with singularities like for piecewise smooth maps with a square root singularity \cite{NORDMARK91}.  Motivated, we next dive deeper in order to understand the reasoning behind such a transition \textcolor{black}{in a smooth three-dimensional discrete Lotka-Volterra map}. \textcolor{black}{Observe that with increase in $R$, the system transits to hyperchaos with two positive Lyapunov exponent, see Fig. \ref{fig:CollageOneParam_Hyperchaos} (b). To understand the role played by the saddle periodic orbits, we continuate the saddle periodic orbits using the multi-dimensional Newton-Raphson method. In Fig. \ref{fig:HyperchaosVaryR} (a) - (b), the saddle period-one orbit is shown in orange, the saddle period-two orbit in pink, the saddle period-six orbit in red, and the stable period-six orbit in blue.  The stable and saddle period-six orbit undergo a saddle-node bifurcation with increase in parameter $R$ at $R=1.1784$.}

\begin{figure*}[!htbp]
\centering
\includegraphics[width=0.9\textwidth]{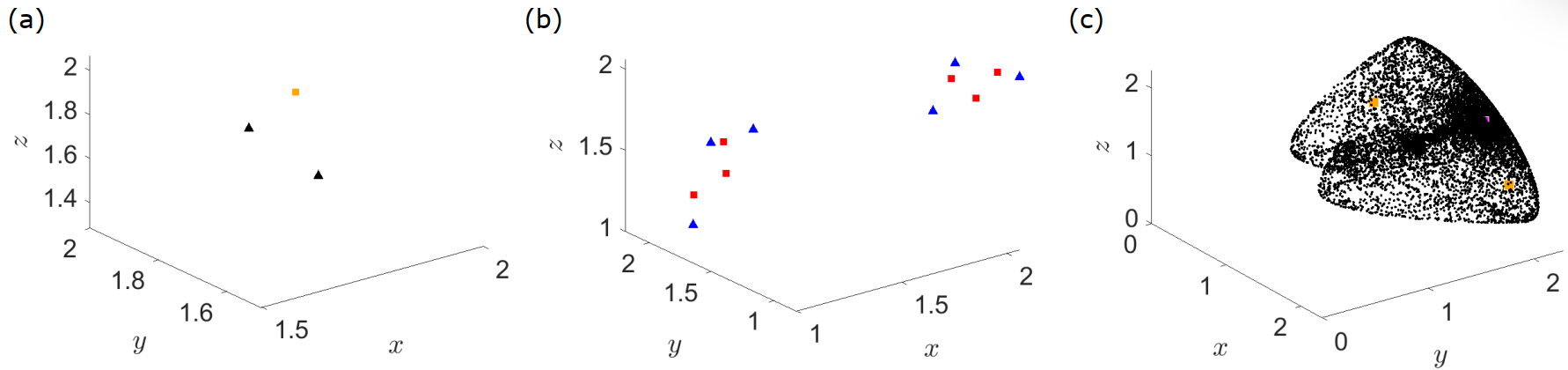}
\caption{Phase space of slices of one-parameter diagram. In (a), for $R=1$,  the coexisting stable period-2 orbit along with the saddle period-1 orbit is shown. In (b), for $R=1.13$, coexisting stable and saddle period-six orbit is \textcolor{black}{shown} in blue and respectively. In (c), for $R=1.2$, hyperchaotic attractor along with absorbed saddle period-2 orbit and saddle period-1 orbit is shown.}
\label{fig:CollagePhaseSpaceThreeLV}
\end{figure*}

\begin{figure}[!htbp]
\centering
\includegraphics[width=0.5\textwidth]{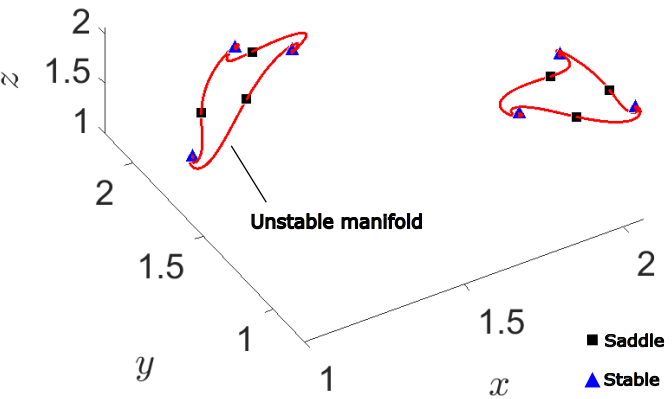}
\caption{Saddle-focus connection between stable and saddle mode-locked period-six orbit.}
\label{fig:SaddleFocusConnection}
\end{figure}

\textcolor{black}{A phase space plot for selected slices are shown in Fig. \ref{fig:CollagePhaseSpaceThreeLV}. In (a), the coexistence of a stable period-two and saddle period-one orbit is shown for $R= 1$. In (b), coexistence of both stable and saddle period-six orbit are shown in blue and red triangles respectively for $R=1.13$. In (c), absorption of saddle period-two and saddle period-one orbit in the hyperchaotic attractor is shown for $R=1.2$. We should emphasize that there is more information associated with Fig. \ref{fig:CollagePhaseSpaceThreeLV} (b). We have observed that the eigenvalues of the saddle period-six orbit has no complex eigenvalues and the eigenvalues of the stable period-six orbit has complex eigenvalues with modulus less than one. Thus, considering the one-dimensional unstable manifolds of the saddle period-six orbit, we get a saddle-focus connection on the disjoint cyclic period-six orbit, see Fig. \ref{fig:SaddleFocusConnection}.}

\begin{figure}[!htbp]
\centering
\includegraphics[width=0.9\textwidth]{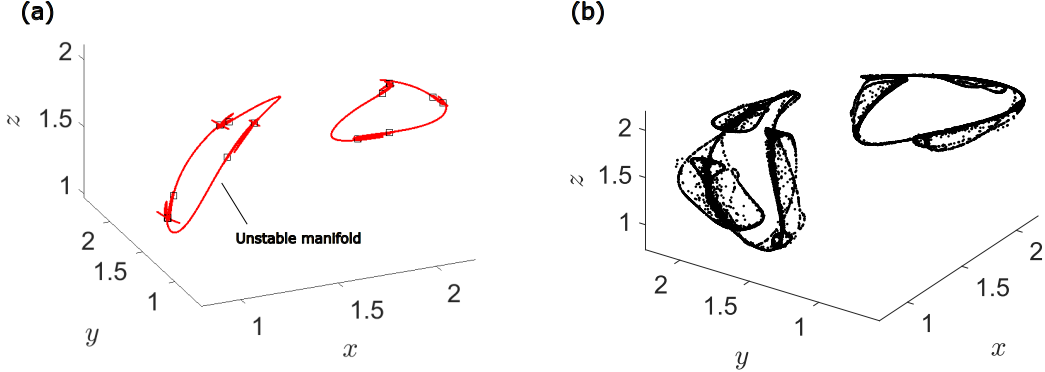}
\caption{A distorted saddle-focus connection. In (a), a distorted saddle-focus connection betweent he saddle period-six orbit at $R=1.152$. For $R=1.17$, a chaotic attractor develops. }
\label{fig:DisortedSf}
\end{figure}

\textcolor{black}{Moreover, as $R$ increases, we observe some interesting transitions in the phase space, see Fig. \ref{fig:DisortedSf}. For $R=1.152$, a distorted saddle-focus connection is observed, see Fig. \ref{fig:DisortedSf} (a) as compared to a regular saddle-focus connection. With increase in $R$ to $1.17$, a chaotic attractor (one positive Lyapunov exponent) develops with an interesting topology. We observe some relevance between between Fig. \ref{fig:DisortedSf} (a) and (b) and it would be interesting to study in detail the mechanism behind this transition and the role played by the one-dimensional unstable manifolds.} 

\begin{figure}[!htbp]
\centering
\includegraphics[width=0.5\textwidth]{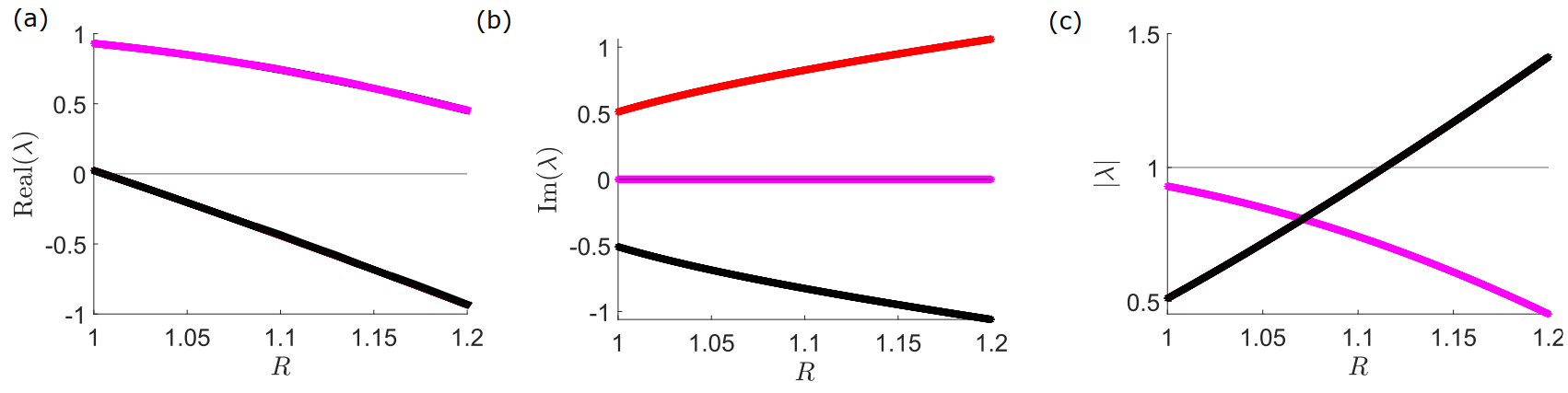}
\caption{Eigenvalue variation of period-two orbit with respect to parameter $R$. In (a), the real part of eigenvalue is shown. In (b), the imaginary part of the eigenvalue, and in (c), the magnitude of the eigenvalue is shown.}
\label{fig:P2EigenvalueThreeLV}
\end{figure}

We analyse the eigenvalues of the period-two orbit so that we can get an insight on the type of bifurcation it underwent \textcolor{black}{so as to bifurcate to a period-six orbit}. A one parameter variation of the eigenvalues with respect to parameter $\beta$ is shown in Fig. \ref{fig:P2EigenvalueThreeLV} (a) - (c). Observe that the eigenvalues are complex and have both non-zero real and imaginary parts. Observe that stable period-two orbit ceases to exist at $R= 1.1104$, see Fig. \ref{fig:CollageOneParam_Hyperchaos} and from Fig. \ref{fig:P2EigenvalueThreeLV} (c), at that parameter value the magnitude of eigenvalue cross unity. Hence, we can conclude that the period-two orbit undergoes a sub-critical Neimark-Sacker bifurcation. This means that an unstable quasiperiodic closed invariant curve is formed. Next, with a small increase in parameter $R$, the unstable closed invariant curve undergoes a saddle-node bifurcation at $R=1.1112$ leading to the formation of a period-six orbit. To understand this event, we compute the saddle period-six orbit and continuate them by the multidimensional Newton-Raphson method. \textcolor{black}{Both the saddle period-one and saddle period-two orbit are continuated in the regime of hyperchaos and we can see that they have two-dimensional unstable manifold and one-dimensional stable manifold indicating the spread in two directions, see Fig. \ref{fig:P2EigenvalueThreeLV}.}

\begin{figure}[!htbp]
\centering
\includegraphics[width=0.5\textwidth]{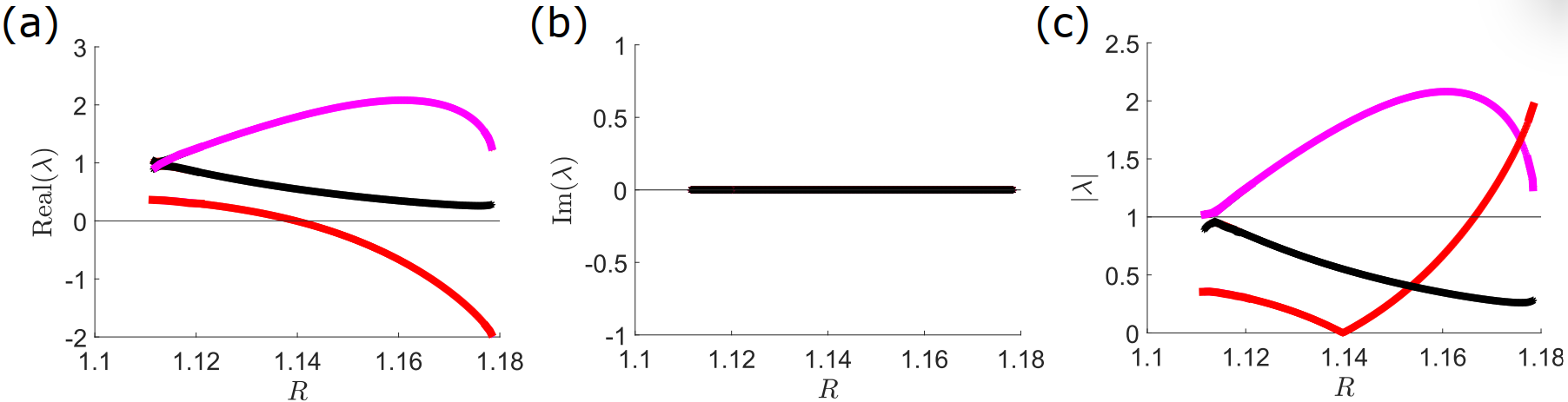}
\caption{Eigenvalue variation of saddle period-six orbit with respect to parameter $R$. In (a), the real part of eigenvalue is shown. In (b), the imaginary part of the eigenvalue, and in (c), the magnitude of the eigenvalue is shown. Observe the occurence of saddle-node bifurcation.}
\label{fig:P6Saddle-EigenvalueThreeLV}
\end{figure}

\begin{figure}[!htbp]
\centering
\includegraphics[width=0.5\textwidth]{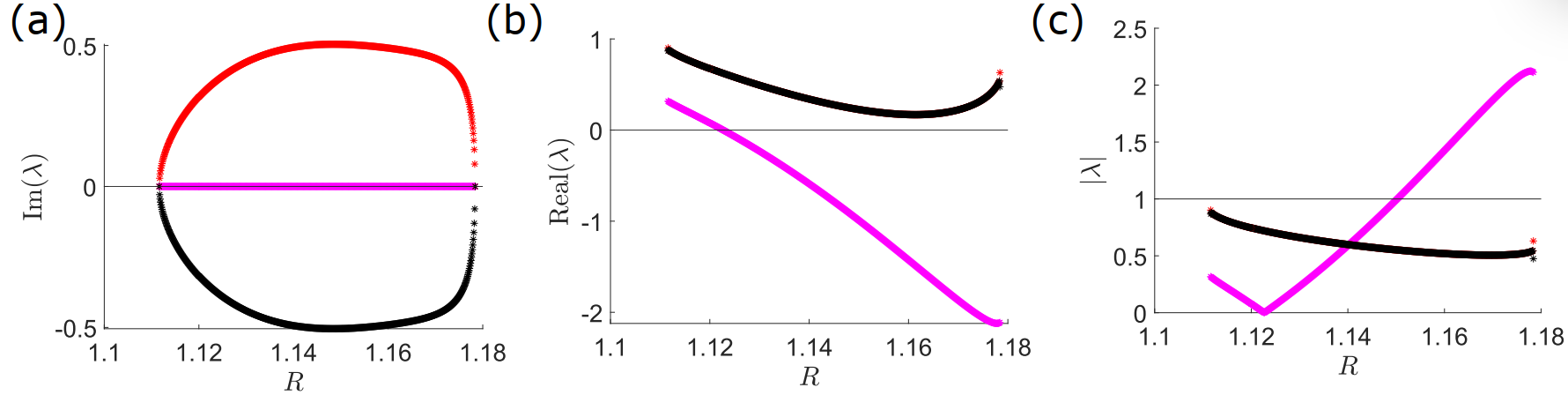}
\caption{Eigenvalue variation of stable period-six orbit with respect to parameter $R$. In (a), the real part of eigenvalue is shown. In (b), the imaginary part of the eigenvalue, and in (c), the magnitude of the eigenvalue is shown. Observe the occurrence of saddle-node bifurcation.}
\label{fig:P6Stable-EigenvalueThreeLV}
\end{figure}

Note that indeed the period-six orbit undergoes a saddle-node bifurcation as the stable period-six orbit (in blue) and saddle period-six orbit (in red) collide and annihilate. The corresponding one-parameter eigenvalue plots of both saddle and stable period-six orbit also illustrate this, see Fig. \ref{fig:P6Saddle-EigenvalueThreeLV}, Fig. \ref{fig:P6Stable-EigenvalueThreeLV} respectively.  For the saddle period-six orbit in Fig. \ref{fig:P6Saddle-EigenvalueThreeLV} (a)-(c), all the \textcolor{black}{eigenvalues} are real and the magnitude of the eigenvalues cross $1$ at $R=1.1112$, showing the \textcolor{black}{occurrence} of a saddle-node bifurcation. Same can also be observed in the case of the eigenvalues of stable period-six orbit in Fig. \ref{fig:P6Stable-EigenvalueThreeLV} (a)-(c).

In the next section, some  bifurcations of quasiperiodic orbits are shown with an interplay between the saddle-node and saddle-focus connections. Some open problems are also illustrated. 

\section{Formation of cyclic closed invariant curves}
\label{sec:cyclic}
\begin{figure*}[!htbp]
\centering
\includegraphics[width=0.5\textwidth]{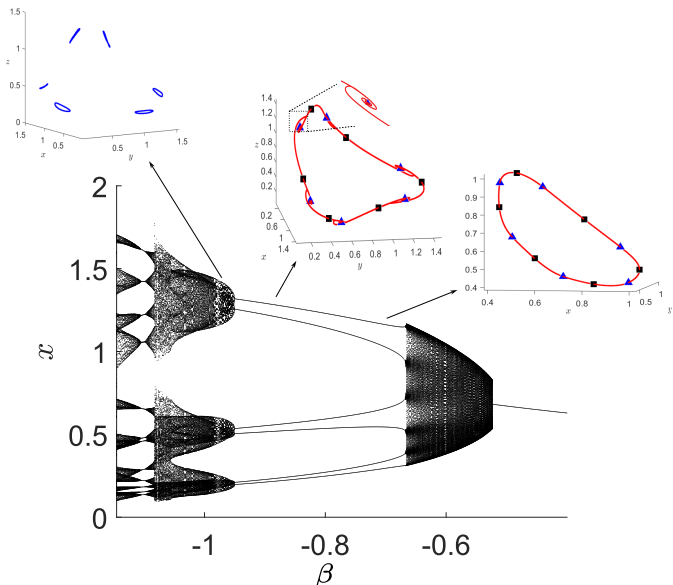}
\caption{A one-parameter bifurcation diagram of $x$ vs $\beta$. Routes to the formation of disjoint cyclic closed invariant curve is discussed.}
\label{fig:snsf}
\end{figure*}

Consider a one-parameter bifurcation of $x$ vs $\beta$ shown in Fig. \ref{fig:snsf}. With decrease in parameter $\beta < -0.6$, we encounter birth of a stable period-six orbit and a saddle period-six orbit via a saddle-node bifurcation. If we consider the unstable manifolds of the saddle period-six orbit near to the saddle-node bifurcation, then we observe a saddle-node connection shown on the rightmost side. Next, as parameter $\beta$ further decreases, the eigenvalues of the stable period-six orbit turns complex and thus when the unstable manifolds of the saddle are considered, it forms a saddle-focus connection instead as the unstable manifold spirals around the stable period-six points. It is well known that the saddle-node connection and saddle-focus connection are not homeomorphic to each other.  With further decrease in $\beta$, the complex eigenvalues of the stable period-six orbit crosses the unit circle and they undergo a supercritical Neimark-Sacker bifurcation leading to the formation of a six cyclic quasiperiodic closed invariant curves.

\section{Conclusions}
\textcolor{black}{A two-parameter Lyapunov chart revealed symmetric bifurcation structure whcih is a promising future research direction to explore. In In this paper, we have shown some additional exotic dynamics in the case of a three-dimensional discrete Lotka-Volterra map. The existence of a three-frequency quasiperiodic torus was observed concerning variation of parameter $\beta$. We have shown the formation of a three-frequency torus via a quasiperiodic-Hopf bifurcation of a closed invariant curve. We then illustrated the existence of quasiperiodic torus doubling bifurcation resulting in the formation of a length-doubled quasiperiodic closed invariant curve. After two subsequent torus doubling bifurcations, the closed invariant curve loses smoothness and the formation of hyperchaotic attractors takes place with two positive Lyapunov exponents. We also show that the saddle fixed point absorbed in the hyperchaotic attractor has two unstable directions. Further, we showcase a peculiar transition from a period-two orbit to a period-six orbit and discuss the underlying mechanism behind their transition. In summary, a period-two orbit undergoes a subcritical Neimark-Sacker bifurcation, and the formation of an unstable closed invariant curve undergoes a saddle-node bifurcation leading to the formation of a period-six orbit. This process was illustrated by continuating the saddle periodic orbit and their corresponding eigenvalues by the multidimensional Newton-Raphson method. We have shown that the structure associated with the mode-locked period-six orbit is that of a saddle-focus connection. Moreover, we have shown a  route to the formation of a disjoint cyclic six loop. It follows the route of saddle-node connection $\rightarrow$ saddle-focus connection $\rightarrow$ disjoint cyclic closed curve. This motivates to ask further: can there be a direct transition of a saddle-node connection to disjoint cyclic closed curves without further transiting to a saddle-focus connection? It would be interesting to further consider a network of Lotka-Volterra map in various configurations like ring-star \cite{Muni20,SolitaryVdP}, lattice \cite{ShMu20a}, multi-layer \cite{ShMu21a} and discuss the prevalence of various spatiotemporal patterns associated.}

\section*{Acknowledgements}
S.S.M acknowledges fruitful discussions with David Simpson during the course of this work.
\section*{Conflict of interest}
 The authors declare that they have no conflict of interest.

 \section*{Data Availability Statement}
 The data that support the findings of this study are available within the article.

\bibliographystyle{unsrt}
\bibliography{Arxiv}

\begin{thebibliography}{10}

\bibitem{May1976}
Robert~M. May.
\newblock Simple mathematical models with very complicated dynamics.
\newblock {\em Nature}, 261(5560):459–467, June 1976.

\bibitem{Kawano2020}
Tomonori Kawano, Nigel Wallbridge, and Carrol Plummer.
\newblock Logistic models for simulating the growth of plants by defining the maximum plant size as the limit of information flow.
\newblock {\em Plant Signal. Behav.}, 15(2):1709718, January 2020.

\bibitem{BravodelaParra2013}
R.~Bravo de~la Parra, M.~Marvá, E.~Sánchez, and L.~Sanz.
\newblock Reduction of discrete dynamical systems with applications to dynamics population models.
\newblock {\em Math. Model. Nat. Phenom.}, 8(6):107–129, 2013.

\bibitem{Bischi2010}
G.I. Bischi and F.~Tramontana.
\newblock Three-dimensional discrete-time {L}otka–{V}olterra models with an application to industrial clusters.
\newblock {\em Commun. Nonlinear Sci. Numer. Simul.}, 15(10):3000–3014, October 2010.

\bibitem{Blackmore2001}
Denis Blackmore, Jerry Chen, John Perez, and Michelle Savescu.
\newblock Dynamical properties of discrete {L}otka–{V}olterra equations.
\newblock {\em Chaos Solitons Fract.}, 12(13):2553–2568, October 2001.

\bibitem{ChChaoui2022}
Mohamed Ch-Chaoui and Karima Mokni.
\newblock A discrete evolutionary {B}everton–{H}olt population model.
\newblock {\em Int. J. Dyn. Control.}, 11(3):1060–1075, September 2022.

\bibitem{Muni23a}
S.S. Muni.
\newblock Mode-locked orbits, doubling of invariant curves in discrete {H}indmarsh-{R}ose neuron model.
\newblock {\em Phys. Scr.}, 98(8), 2023.

\bibitem{ZS22AEU}
Z.T. Njitacke, T.F. Fozin, S.S. Muni, J.~Awrejcewicz, and J.~Kengne.
\newblock Energy computation, infinitely coexisting patterns and their control from a {H}indmarsh-{R}ose neuron with memristive autapse: Circuit implementation.
\newblock {\em AEU Int. J. Electron. Commun.}, 155:154361, 2022.

\bibitem{ZSChaos22}
Z.T. Njitacke, S.S. Muni, T.F. Fonzin, G.D. Leutcho, and J.~Awrejcewicz.
\newblock Coexistence of infinitely many patterns and their control in heterogeneous coupled neurons through a multistable memristive synapse.
\newblock {\em Chaos}, 32(5):053114, 2022.

\bibitem{Chialvo22}
S.S. Muni, H.O. Fatoyinbo, and I.~Ghosh.
\newblock Dynamical effects of electromagnetic flux on {C}hialvo neuron map: Nodal and network behaviors.
\newblock {\em Int. J. Bifurc. Chaos.}, 32(09):2230020, 2022.

\bibitem{IZH22}
S.S. Muni, K.~Rajagopal, A.~Karthikeyan, and S.~Arun.
\newblock Discrete hybrid {I}zhikevich neuron model: Nodal and network behaviours considering electromagnetic flux coupling.
\newblock {\em Chaos Solitons Fract.}, 155:111759, 2022.

\bibitem{Chalmers2015}
Alexander~D. Chalmers, Anna Cohen, Christina~A. Bursill, and Mary~R. Myerscough.
\newblock Bifurcation and dynamics in a mathematical model of early atherosclerosis: How acute inflammation drives lesion development.
\newblock {\em J. Math. Biol.}, 71(6–7):1451–1480, March 2015.

\bibitem{Mukherjee2024}
Debasmita Mukherjee, Sishu~Shankar Muni, and Hammed~Olawale Fatoyinbo.
\newblock A dynamical system framework for a mathematical model of atherosclerosis.
\newblock {\em Franklin Open}, 7:100116, May 2024.

\bibitem{Adachi2022}
Kyosuke Adachi, Ryosuke Iritani, and Ryusuke Hamazaki.
\newblock Universal constraint on nonlinear population dynamics.
\newblock {\em Commun. Phys.}, 5(1):2399--3650, 2022.

\bibitem{Muni2024}
Sishu~Shankar Muni.
\newblock Ergodic and resonant torus doubling bifurcation in a three-dimensional quadratic map.
\newblock {\em Nonlinear Dyn.}, 112:4651–4661, January 2024.

\bibitem{Vibr15}
T~Bakri, Y~A Kuznetsov, and F~Verhulst.
\newblock Torus bifurcations in a mechanical system.
\newblock {\em J. Dyn. Differ. Equ.}, 27:371--403, 2015.

\bibitem{Ma2023}
Jiying Ma and Mingxia Duan.
\newblock Codimension-two bifurcations of a two-dimensional discrete time {L}otka-{V}olterra predator-prey model.
\newblock {\em Discrete Contin. Dyn. Syst. - B}, 29(3):1217–1242, 2024.

\bibitem{Balibrea2006}
Francisco Balibrea, Juan Luis~García Guirao, Marek Lampart, and Jaume Llibre.
\newblock Dynamics of a {L}otka–{V}olterra map.
\newblock {\em Fund. Math.}, 191(3):265–279, 2006.

\bibitem{LuisGarcaGuirao2008}
Juan Luis García~Guirao and Marek Lampart.
\newblock Transitivity of a {L}otka-{V}olterra map.
\newblock {\em Discrete Contin. Dyn. Syst. - B}, 9(1):75–82, 2008.

\bibitem{Yang2020}
Feifei Yang, Jun Mou, Jian Liu, Chenguang Ma, and Huizhen Yan.
\newblock Characteristic analysis of the fractional-order hyperchaotic complex system and its image encryption application.
\newblock {\em Signal Process.}, 169:107373, April 2020.

\bibitem{KOCAREV1992}
Lj. Kocarev, K.~S. Halle, K.~Eckert, L.~O. Chua, and U.~Parlitz.
\newblock Experimental demonstration of secure communications via chaotic synchronization.
\newblock {\em International Journal of Bifurcation and Chaos}, 2(03):709–713, September 1992.

\bibitem{Guchi16}
Yutaka Shimada, Emiko Takagi, and Tohru Ikeguchi.
\newblock Symmetry of {L}yapunov exponents in bifurcation structures of one-dimensional maps.
\newblock {\em Chaos}, 26(12):1089--7682, December 2016.

\bibitem{NORDMARK91}
A.B. Nordmark.
\newblock Non-periodic motion caused by grazing incidence in an impact oscillator.
\newblock {\em J. Sound Vib.}, 145(2):279--297, 1991.

\bibitem{Muni20}
S.~S. Muni and A.~Provata.
\newblock Chimera states in ring–star network of {C}hua circuits.
\newblock {\em Nonlinear Dyn.}, 101:2509–2521, 2020.

\bibitem{SolitaryVdP}
V.~{Santos}, M.~Rolim Sales, S.~S. Muni, J.~D. Szezech, A.~M. Batista, S.~Yanchuk, and J.~Kurths.
\newblock Identification of single- and double-well coherence–incoherence patterns by the binary distance matrix.
\newblock {\em Commun. Nonlinear Sci. Numer. Simul.}, 125:107390, 2023.

\bibitem{ShMu20a}
I.~A. Shepelev, A.~V. Bukh, S.~S. Muni, and V.~S. Anishchenko.
\newblock Role of solitary states in forming spatiotemporal patterns in a 2d lattice of {V}an der {P}ol oscillators.
\newblock {\em Chaos Solitons Fract.}, 135:109725, 2020.

\bibitem{ShMu21a}
I.~A. Shepelev, S.~S. Muni, E.~Schöll, and G.~I. Strelkova.
\newblock Repulsive inter-layer coupling induces anti-phase synchronization.
\newblock {\em Chaos}, 31:063116, 2021.

\end{thebibliography}

\end{document}